\documentclass[a4paper]{article}
\usepackage{amscd,amsmath,amssymb}

\newcommand{\p}[1]{(\ref{#1})}

\newcommand{\cH}{{\cal H}}

\newcommand{\cN}{{\cal N}}
\newcommand{\cQ}{{\cal Q}}

\newcommand{\cbQ}{\overline{\cal Q}}

\newcommand{\cA}{{\cal A}}

\newcommand{\bQ}{{\overline Q}{}}

\newcommand{\bxi}{{\bar\xi}}

\newcommand{\bpsi}{{\bar\psi}{}}

\newcommand{\brho}{{\bar\rho}{}}

\newcommand{\und}{\qquad\textrm{and}\qquad}
\renewcommand{\=}{\ =\ }

\newcommand{\be}{\begin{equation}}
\newcommand{\ee}{\end{equation}}
\newcommand{\bea}{\begin{eqnarray}}
\newcommand{\eea}{\end{eqnarray}}

\newcommand{\ba}{\begin{array}} \newcommand{\ea}{\end{array}}

\def\im{{\rm i}}
\def\sfrac#1#2{{\textstyle\frac{#1}{#2}}}

\newcommand{\nn}{\nonumber}

\begin{document}
\phantom{.}
\vspace{3cm}

\begin{center}
{\LARGE\bf  On $\cN{=}\,2$ supersymmetric Ruijsenaars--Schneider models  }
\end{center}
\vspace{1cm}

\begin{center}
{\Large\bf  Sergey~Krivonos${}^a$ \ and \ Olaf~Lechtenfeld${}^b$}
\end{center}

\vspace{0.2cm}

\begin{center}

{${}^a$ \it
Bogoliubov  Laboratory of Theoretical Physics, JINR,
141980 Dubna, Russia}

${}^b$ {\it
Institut f\"ur Theoretische Physik and Riemann Center for Geometry and Physics \\
Leibniz Universit\"at Hannover,
Appelstrasse 2, 30167 Hannover, Germany}

\vspace{0.5cm}

{\tt e-mails: krivonos@theor.jinr.ru, olaf.lechtenfeld@itp.uni-hannover.de}
\end{center}
\vspace{2cm}

\begin{abstract}
\noindent 
We construct an $\cN{=}\,2$ supersymmetric extension of $n$-particle Ruijsenaars--Schneider models. 
The guiding feature is a deformation of the phase space. The supercharges have a ``free" form 
linear in the fermions but produce an interacting four-fermion Hamiltonian. A field-dependent unitary 
transformation maps to standard fermions obeying conventional Poisson brackets. In this frame,
the supercharges and Hamiltonian have long ``fermionic tails''. We also comment on previous attempts 
in this direction.
\end{abstract}


\vskip 0.5cm


\thispagestyle{empty}

\newpage
\setcounter{page}{1} 
\setcounter{equation}{0}
\section{Introduction}
The Ruijsenaars--Schneider models~\cite{RS1}  are known for more than three decades 
as a ``relativistic" variant\footnote{
For a justification of the term "relativistic", see e.g.~the discussion in \cite{BS}.} 
of the well known Calogero--Moser--Sutherland  models~\cite{Calogero,Calogero1,Poly1} . 
An $\cN{=}\,2$ supersymmetric extension of the latter has been constructed many years ago~\cite{susyCal, susyCal1}. 
Soon after, its symmetry algebra and eigenfunctions have been analyzed~\cite{susyCal2}, 
introducing the Jack superpolynomials~\cite{susyCal3} as superspace analogs of the Jack polynomials. 
In contrast, its``relativistic" cousin -- an $\cN{=}\,2$ supersymmetric extension of Ruijsenaars--Schneider models
-- remained almost completely unexplored for a couple of decades. Presumably responsible for this relative silence
is the unfamiliar structure of its action (the lack of a potential), which impedes trusted techniques of
supersymmetric mechanics~\cite{FM}  in a relativistic setting.

The first (integrable) $\cN{=}\,2$ supersymmetric generalization of the (quantum) trigonometric 
Ruijsenaars--Schneider model has been reported in~\cite{SRS}. Its construction is based on
(a) the integrability of the bosonic system, 
(b) a modification of the anticommutation relations between fermions, and
(c) a complicated definition of adjoints.
We shall comment on it in the Conclusions.

A second example of an $\cN{=}\,2$ supersymmetric Ruijsenaars--Schneider system was elaborated in \cite{ag}
for the case of three particles. Its ansatz for the supercharges mimics those in the supersymmetric
Calogero--Moser--Sutherland models~\cite{susyCMS1, susyCMS2, susyCMS3}. 
Unfortunately, it is unclear how this permutation-symmetry breaking solution can be extended 
to an arbitrary number of particles. 

Here, we succeed in constructing $\cN{=}\,2$ supersymmetric $n$-particle Ruijsenaars--Schneider  models. 
Our guidelines are:
\begin{itemize}
\item The supercharges $\cQ$ and $\cbQ$ are taken to be ``free'', i.e.~only linear in the fermions,
\item The``interactions" are entirely encoded in a highly non-trivial structure of the Poisson brackets. 
\end{itemize} 
With this prescription, $\cN{=}\,2$ supercharges may easily be constructed. 
The $\cN{=}\,2$ {Poincar\'{e} superalgebra then produces an interacting Hamiltonian.
With a (field-dependent) unitary transformation one comes back to the standard fermions
with standard Poisson brackets. The ensuing supercharges appear as a natural generalization 
of the ``non-relativistic" fermions to the ``relativistic" case. 
When passing to the standard fermions, the Hamiltonian gets modified by a long ``fermionic tail''.

The paper is organized as follows. 
In Section~2 we briefly review the relevant properties of the bosonic Ruijsenaars--Schneider model, 
emphasizing on its free-Hamiltonian representation, whose price is deformed Poisson brackets. 
Section~3 extends the non-standard bosonic Poisson brackets to the $\cN{=}\,2$ supersymmetric case,
where the ``free" supercharges produce a proper interacting $\cN{=}\,2$ supersymmetric Hamiltonian. 
The precise relation with standard fermions (obeying standard Poisson brackets) is presented in Section~4,
which also computes the Hamiltonian in this frame.
In the Conclusions we compare with the previous efforts \cite{SRS,ag} 
and mention possible further developments.

\setcounter{equation}{0}
\section{Bosonic models}
The Ruijsenaars--Schneider models are integrable many-body systems in one dimension which are described by the equations of motion \cite{RS1}
\be\label{ieq1}
\ddot{x}_i \= 2 \sum^n_{j\neq i} \dot{x}_i \dot{x}_j W(x_i{-}x_j)\ ,
\ee
where the function $W$ is one of the following functions\footnote{
We will not consider the elliptic variant in this paper.}
\be\label{W}
W(x)\ \in\ \bigl\{ 1/x, \ 1/\sin(x), \ 1/\sinh(x), \ 1/\tan(x), \ 1/\tanh(x) \bigr\} \ .
\ee 
The standard description of such systems is based on the Hamiltonian \cite{RS1}
\be\label{S1}
H \=\sfrac{1}{2} \sum^n_i e^{2\theta_i} \prod^n_{j (\neq i)} f(x_i{-}x_j)\ ,
\ee 
where the rapidities $\theta_i$ and the coordinates $x_j$ obey standard Poisson brackets,
\be\label{PB0}
\left\{ x_i , \theta_j \right\} =\delta_{ij} \und \left\{x_i,x_j \right\} = \left\{ \theta_i, \theta_j\right\} =0\ .
\ee
The functions $W$ in \p{ieq1} and $f$ in \p{S1} are related (in this order):
\bea\label{fW}
f(z) \,\in\,\Bigl\{ \frac{1}{z} , \frac{1}{\sinh(z)}, \frac{1}{\sin(z)}, \frac{1}{\tanh(\frac{z}{2})}, \frac{1}{\tan(\frac{z}{2})} \Bigr\}
\qquad\Leftrightarrow\qquad
W(z)\,\in\,\Bigl\{ \frac{1}{z}, \frac{1}{\tanh(z)}, \frac{1}{\tan(z)}, \frac{1}{\sinh(z)}, \frac{1}{\sin(z)} \Bigr\}\ .
\eea

We prefer the following re-interpretation of the Ruijsenaars--Schneider  systems. 
Let us cast the Hamiltonian $S_{+1}$ into a free form,\footnote{
This form of the Hamiltonians and Poisson brackets is explicitly written in \cite{ag} but may be older.}
\be\label{Hb}
H \= \sfrac{1}{2} \sum_{i=1}^n p_i^2\ ,
\ee
with new momenta
\be\label{p}
p_i \= e^{\theta_i} \prod^n_{j (\neq i)} \sqrt{f(x_i{-}x_j)}\ .
\ee
This redefinition ($\theta_i\to p_i$) clearly changes the Poisson brackets to\footnote{
We assume that $W(0)=0$, which allows us to avoid writing
$ \left\{ p_i, p_j\right\}=\left(1-\delta_{ij}\right) p_i p_j W(x_i-x_j)$.}
\be\label{case1}
\left\{ x_i, p_j\right\} = \delta_{ij} p_j \und \left\{ p_i, p_j\right\}=p_i p_j W(x_i{-}x_j)\ .
\ee 

One may check that the Hamiltonian $H$ \p{Hb} and brackets \p{case1} result in the equations of motion \p{ieq1}. Indeed, from   \p{Hb} and \p{case1} we have
\be\label{eom1a}
\dot{x}_i \ \equiv\  \left\{ x_i, H\right\} \= p_i^2\ ,
\ee
and, therefore,
\be\label{eom1b}
\ddot{x}_i \ \equiv\  \left\{ \dot{x}_i, H\right\} \= \left\{ p_i^2 ,H\right\} 
\= 2 \sum_{j (\neq i)}^n p_i^2 p_j^2 W(x_i{-}x_j) \= 2  \sum_{j (\neq i)}^n \dot{x}_i \dot{x} _j W(x_{i}{-}x_j) \ ,
 \ee
 as it should be.

\setcounter{equation}{0}
\section{ $\cN{=}\,2$ supersymmetric Ruijsenaars--Schneider models}
We have seen that the bosonic Ruijsenaars--Schneider  models can be described by  a free Hamiltonian, 
while the interaction moved to the Poisson brackets. We shall use the same strategy to construct 
an $\cN{=}\,2$ supersymmetric extension of the Ruijsenaars--Schneider models.

Such a model is equivalent to the existence of supercharges
 $\cQ$ and $ \cbQ$ forming an $\cN{=}\,2$ Poincar\'{e} superalgebra
\be\label{N2sa}
\left\{ \cQ, \cbQ \right\} = - 2 \im \cH \und \left\{ \cQ, \cQ \right\}=\left\{ \cbQ, \cbQ \right\}=0
\ee
together with the Hamiltonian $\cH$ whose bosonic sector coincides with the Hamiltonian $H$ \p{Hb}.

To construct such supercharges we extend the  $2\,n$ phase-space variables $x_i$ and $p_j$, 
obeying the brackets \p{case1}, by  $2\,n$  fermions $\psi_i$ and $\bpsi_{j}=\left(\psi_j\right)^\dagger$ , 
subject to the brackets
\bea\label{PB}
&& \left\{ \psi_i, \psi_j \right\} \= -  \psi_i \psi_j W(x_i{-}x_j)\ , \quad
\left\{ \bpsi_i, \bpsi_j \right\} \= -  \bpsi_i \bpsi_j W(x_i{-}x_j)\ , \quad 
\left\{ \psi_i, \bpsi_{j}\right\} \= -\im\, \delta_{ij} +\psi_i \bpsi_j W(x_i{-}x_j)\ ,\nn\\[10pt]
&& \left\{p_i , \psi_j \right\} \= \ \sfrac{\im}{2} \delta_{ij} \, p_i \psi_i\, \sum_k \psi_k \bpsi_k W'(x_i{-}x_k) 
- \sfrac{\im}{2} p_i \psi_j \psi_i \bpsi_i W'(x_i{-}x_j)\ , \qquad \left\{ x_i , \psi_j \right\}=0\ , \\
&& \left\{p_i , \bpsi_j \right\} \ = -\sfrac{\im}{2} \delta_{ij}\,p_i \bpsi_i \, \sum_k \psi_k \bpsi_k W'(x_i{-}x_k) 
+ \sfrac{\im}{2} p_i \bpsi_j \psi_i \bpsi_i W'(x_i{-}x_j)\, \qquad \left\{ x_i , \bpsi_j \right\}=0 \ .\nn
\eea
It is rather easy to check that the Jacobi identities are fulfilled.

Finally, we verify that the supercharges
\be\label{QQb1}
\cQ= \sum_{i}^n p_i \psi_i \und
\cbQ= \sum_{i}^n p_i  \bpsi_i
\ee
form an $\cN{=}\,2$ Poincar\'{e} superalgebra \p{N2sa} with the Hamiltonian
\be\label{H1}
\cH\= \sfrac{1}{2} \sum_{i=1}^n p^2_i 
+\im \sum_{i,j}^n p_i p_j \psi_i \bpsi_j W(x_i{-}x_j)
+ \sfrac{1}{2} \sum_{i,j}^n p^2_i \psi_i \bpsi_i \psi_j\bpsi_j W'(x_i{-}x_j)\ . 
\ee
Thus, the set $\{\cQ, \cbQ,\cH\}$ describes an $\cN{=}\,2$ supersymmetric  extension 
of the  Ruijsenaars--Schneider  models.

It should be noted that everything works fine for {\it any\/} antisymmetric function $W(x)$. 
Thus, the explicit choices in \p{W} are dictated by integrability and not by $\cN{=}\,2$ supersymmetry.

Finally, we comment on the rationale leading to the form of the brackets \p{PB}.
We insist that the supercharges $Q$ and $\bQ$ obey the brackets $\left\{Q,Q\right\} = \left\{\bQ,\bQ\right\}=0$ 
and have the  structure \p{QQb1}. To achieve this, the brackets 
$\left\{ \psi_i, \psi_j \right\} = -  \psi_i \psi_j W(x_i{-}x_j)$ and
$\left\{ \bpsi_i, \bpsi_j \right\} = -  \bpsi_i \bpsi_j W(x_i{-}x_j)$
compensate contributions coming from $\left\{p_i, p_j\right\}$.
For a complete cancellation of $\left\{ Q,Q \right\}$, we further make the ansatz
$\left\{ p_i, \psi_j\right\} = p_i \psi_i {\cA_{i,j}}$.
The functions ${\cA_{i,j}}$ as well as the yet undetermined brackets $\left\{\psi_i, \bpsi_j\right\}$ 
finally follow uniquely from the Jacobi identities.

\setcounter{equation}{0}
\section{Playing with the fermions}
One-dimensional supersymmetric systems feature a rich possibility to define the fermions.  For example,
the quite nonlinear redefinition of the fermions in \cite{KLPS} brings the supercharges of the $\cN$-extended supersymmetric $A_{n}$ Calogero model introduced in \cite{KLS} to the standard form maximally cubic in the fermions. The mystery with the brackets \p{PB} has the
same origin. Indeed, one may define alternative fermions
\be\label{xi}
\xi_i \= \exp\Bigl\{ \sfrac{\im}{2}\, \sum_j^n \psi_j \bpsi_j W(x_i{-}x_j)\Bigr\} \ \psi_i \und
\bxi_i \= \exp\Bigl\{ -\frac{\im}{2}\, \sum_j^n \psi_j \bpsi_j W(x_i{-}x_j)\Bigr\} \ \bpsi_i\ ,
\ee
which are subject to the standard brackets
\be\label{PBxi}
\left\{\xi_i, \xi_j\right\} =\left\{\bxi_i, \bxi_j\right\}=0\ ,\quad 
\left\{\xi_i, \bxi_j\right\} =-\im\, \delta_{ij} \und 
\left\{p_i, \xi_j\right\}= \left\{p_i, \bxi_j\right\}=0\ .
\ee
This exponential redefinition is inspired by the relation~\p{p} between momenta and rapidities.
Observing that
\be
\xi_i \bxi_i = \psi_i \bpsi_i \qquad \forall i \ ,
\ee
the inverse transformation reads
\be\label{ixi}
\psi_i \= \exp\Bigl\{- \frac{\im}{2}\, \sum_j^n \xi_j \bxi_j W(x_i{-}x_j)\Bigr\} \ \xi_i,\quad
\bpsi_i \= \exp\Bigl\{ \frac{\im}{2}\, \sum_j^n \xi_j \bxi_j W(x_i{-}x_j)\Bigr\} \ \bxi_i\ .
\ee
Thus, the supercharges \p{QQb1} and the Hamiltonian \p{H1} can be represented as
\bea
\cQ &=& \sum_{i}^n p_i \exp\Bigl\{- \frac{\im}{2}\, \sum_j^n \xi_j \bxi_j W(x_i{-}x_j)\Bigr\}\,\xi_i \ ,\qquad
\cbQ \= \sum_{i}^n p_i  \exp\Bigl\{ \frac{\im}{2}\, \sum_j^n \xi_j \bxi_j W(x_i{-}x_j)\Bigr\}\,\bxi_i\ , 
\label{QQH1} \\[4pt]
\cH&=& \sfrac{1}{2} \sum_{i=1}^n p^2_i \ +\ \im \sum_{i,j}^n p_i p_j 
\exp\Bigl\{ -\frac{\im}{2} \sum_k \xi_k \bxi_k \bigl(W(x_i{-}x_k)-W(x_j{-}x_k)\bigr)\Bigr\}\,
\xi_i \bxi_j W(x_i{-}x_j) \nn \\
&&
\ +\ \sfrac{1}{2} \sum_{i,j}^n p^2_i \xi_i \bxi_i \xi_j\bxi_j W'(x_i{-}x_j). 
\label{QQH2}
\eea
Thus, in this frame, the supercharges $\cQ$ and $\cbQ$ 
and the Hamiltonian $\cH$ acquire long ``fermionic tails''.

\setcounter{equation}{0}
\section{Conclusions}
We have constructed an $\cN{=}\,2$ supersymmetric extension of the Ruijsenaars--Schneider system,
starting from the bosonic equations of motion~\p{ieq1}.
Our construction is valid for any antisymmetric function $W$. 
Only the demand of integrability will restrict this function to the choices known from Ruijsenaars--Schneider  models.
When re-expressed in terms of standard fermions, the Hamiltonian and the supercharges are natural ``relativistic'' generalizations of the ``non-relativistic'' ones.

Let us compare our results with previous attempts on this problem.
The work of \cite{SRS} contains fermionic brackets similar to~\p{PB} between $\psi_i$ and $\bpsi_j$. 
However, it does not feature deformed Poisson brackets between fermions and momenta like in~\p{PB},
but rather introduces complicated conjugation properties.
Of course, our construction is pure classical and, for the time being, ignores integrability,
but we do not foresee obstacles for implementing the latter.

A three-particle Ruijsenaars--Schneider model was proposed in~\cite{ag}. 
Its supercharges 
\bea\label{Gal1}
Q &=& \sum_{i=1}^3 p_i \xi_i  - \im\, p_1 \xi_1 \xi_3 \bxi_3 W(x_1{-}x_3) 
 - \im\, p_2 \xi_2 \xi_1 \bxi_1 W(x_2{-}x_1)  - \im\, p_3 \xi_3 \xi_2 \bxi_2 W(x_3{-}x_2)\ , \nn \\
\bQ &=& \sum_{i=1}^3  p_i \bxi_i + \im\, p_1 \bxi_1 \xi_3 \bxi_3 W(x_1{-}x_3)  
+ \im\, p_2 \bxi_2 \xi_1 \bxi_1 W(x_2{-}x_1)  + \im\, p_3 \bxi_3 \xi_2 \bxi_2 W(x_3{-}x_2)
\eea
indeed form an $\cN{=}\,2$ Poincar\'{e} algebra \p{N2sa}, 
and the fermions $\xi_i$ and $\bxi_j$ obey the standard brackets~\p{PBxi}.
To compare these supercharges to ours one should put them into the form
\be
Q  =  \sum_i^3 p_i \rho_i \und \bQ  = \sum_i^3 p_i \brho_i
\ee
with new fermions
\be\label{rho}
\rho_i \= \xi_i \,\bigl( 1- \im \xi_{i-1} \bxi_{i-1} W(x_i{-}x_{i-1}) \bigr)
\qquad\text{cyclicly in} \quad i=1,2,3\ .
\ee
While the brackets $\left\{\rho_i,\rho_j\right\}$, $\left\{\brho_i,\brho_j\right\}$ and $\left\{\rho_i,\brho_j\right\}$ 
coincide with the corresponding ones for $\psi_i$ and $\bpsi_i$ from~\p{PB}, 
the non-covariant structure of the $\rho$ fermions results in completely different brackets 
$\left\{p_i, \rho_j\right\}$ and $\left\{p_i, \brho_j\right\}$. 
For example, it follows from \p{rho} that $\left\{ p_1, \rho_3\right\}=0$ in explicit contradiction
with  $\left\{ p_1, \psi_3\right\} \neq 0$ from \p{PB}, when restricted to the three-particle case. 
We conclude that our model for $n{=}3$ differs from the one in~\cite{ag}.
Another road to the same conclusion expands the supercharges~\p{QQH1} in powers of fermions. 
Even for three particles, \p{QQH1} produces five-fermion terms, in the contradiction with the ansatz in~\cite{ag}.

There is a number of interesting open issues regarding these extended supersymmetric 
Ruijsenaars--Schneider models.
The list contains prospective integrability, an off-shell superfield Lagrangian formulation, and
$\cN{=}\,4$ generalizations. We plan to clarify some of these points elsewhere.

\vspace{0.5cm}

\noindent{\bf Acknowledgements}

\noindent
We thank Nikolay Kozyrev and Armen Nersessian for  fruitful discussions.
This work was supported by the RFBR-DFG grant No 20-52-12003.
We are grateful to the referee who pointed out a mistake in \p{QQH2} in the first version of this paper, 
where only the first two Taylor terms of the exponential function had been written.

\end{document}